\documentclass[onecolumn,epsfig,pre]{revtex4}
\usepackage{graphicx}
\usepackage{amssymb}
\usepackage{amsmath}
\usepackage{hyperref}
\usepackage{latexsym}

\def\be{\begin{equation}}
\def\ee{\end{equation}}
\def\bea{\begin{eqnarray}}
\def\eea{\end{eqnarray}}

\draft

\begin{document}

\title{Phase segregation and transport in a two species multi-lane system}
\author{Sudipto Muhuri$^{1}$ and Ignacio Pagonabarraga$^{2}$}
\affiliation {$^1$Institute of Physics, Sachivalaya Marg, Bhubaneswar 751005, India\\
$^2$ Departament de Fisica Fonamental, Universitat de Barcelona, C.Marti i Franques 1, 08028 Barcelona, Spain}
\begin{abstract} 
We present a two channel driven lattice gas model with oppositely directed species moving on two parallel lanes with lane switching processes. We study correlated lane switching mechanism for particles so that switching may occur with finite probability only when oppositely directed species meet on the same channel. The system is analyzed for  closed ring with conserved total particle number. For asymmetric particle exchange between the lanes, the system exhibits unique polarization phenomenon with  segregation of oppositely directed species between the two lanes. The polarization phenomenon can be understood as a consequence of existence of an {\it absorbing} steady state. For symmetric exchange rate of particles between the lanes, the system remains unpolarized, with equal particle density on both the lanes in the {\it thermodynamic limit} of large system size. We study the system using a combination of a Mean Field(MF) analysis and Monte Carlo simulations. The nature of phase segregation that we see for this system, is distinct from driven particle systems which are in contact with particle reservoir. The features observed for this minimal model will have ramifications for biofilament based intracellular transport, wherein  cellular cargo, e.g; organelles and vesicles are transported by oppositely directed particles on multiple filament tracks.
\end{abstract}

\maketitle
\section{Introduction}
Driven lattice gas models for transport have been a subject of considerable interest due to their ability to describe, in a simplified framework, a number of biologically motivated processes, such as the motion of ribosomes in m-RNA~\cite{mcdonald}, proton conduction in water channels \cite{chou1} or  motor protein driven vesicle transport~\cite{welte,howard} among others. Driven lattice gases provide a general framework to understand the physical properties of these biologically motivated systems and their interrelation with other driven systems, such as  vehicular traffic~\cite{vehicle}. These driven diffusive systems exhibit a rich phase diagram characterized by novel non-equilibrium steady states with finite macroscopic  current \cite{schutzrev,nonequi2,privman,barma5,somen1,somen2} and boundary induced phase transitions \cite{schutzrev,evans,freylet,santen,kolo,somen1,somen2} which do not exist in equilibrium. A specific example of such models is the Totally Asymmetric Exclusion Process (TASEP), in which the particles hop unidirectionally on one lane with a single rate and interact via hard-core repulsion. For TASEP, the complete non-equilibrium phase diagram in terms of particle fluxes at the boundaries, has been worked out explicitly \cite{evans}. The competition between particle fluxes at the boundaries with the process of particle exchange with the surrounding bath has also been addressed and it has been shown that this competition leads to new phase transitions~\cite{freylet}. However, relatively less attention has been paid to asymmetric transport along multiple channels.\\
Transport involving multiple lanes has been observed in the context of intracellular transport of organelles on microtubules \cite{roop, hollenbeck}. Filament based intracellular transport often involves oppositely directed motors, which use multiple array of cytoskeletal filaments, to transport cellular cargoes such as mitochondria, endosomes, lipid droplets and pigment granules~\cite{welte}.
Since there exist alternative scenarios which can lead to bidirectional active cargo transport, it is pertinent to explore the  different mechanisms by which  efficient transport and regulation of cellular cargo is achieved and jamming avoided. Some of the previous theoretical attempts have focused on combining  stochastic directional switching of the transported cargo  with  TASEP like hopping rules  on a single track, to describe transport on a single filament ~\cite{ignaepl,ignapre,madan}.  This corresponds to a biological situation which involves an interplay between tug-of-war of the transported cargo due to oppositely directed motors and cooperative  effective transport.\\
 In this paper we will focus on the role of lane switching by the transported cargo as an alternative scenario to  induce and tune bidirectional transport. Switching between filaments has been observed  for example in axonal transport of mitochondria in neurons~\cite{welte}. Further, the strict exclusion rules of TASEP does not seem to hold true for cellular transport, so that oppositely directed cellular cargo can pass through each other \cite{roop}. In order to circumvent these drawbacks, we present a two lane model which incorporates bidirectionality and  correlated lane switching processes, wherein oppositely directed particles switch lanes with certain finite probability on encountering each other and they are also allowed to  pass through each other with certain finite probability on the same lane.\\ 
Different mechanisms of multiple-lane driven transport have been proposed and studied theoretically~\cite{pronina1,pronina2, harris, mitsudo, popkov1,popkov2, juhasz1,juhasz2,rolland,frey2lane, frey2lane1,santen2,sugden}. While inter-lane  switching has not been considered in Refs.~\cite{popkov1,popkov2}, stochastic switching between  lanes was explicitly incorporated  in Refs.~\cite{ juhasz1,juhasz2, santen2, pronina1, pronina2,frey2lane,frey2lane1,sugden}. Specifically, we will consider a scenario where isolated motors show a tendency to switch between different lanes when hindered by an oppositely directed particle moving on the same lane. Such hindrances can modify the motor affinity to the filament and thus inducing the motor to switch to the neighbouring lane. Hence we will take into account the intrinsically correlated nature of lane hopping and will consider that a  lane switching event can occur only when oppositely directed species encounter each other on a given lane. Such correlated  switching  events give rise to steady state properties which differ qualitatively  from previously proposed single motor switching mechanisms. In particular, correlated switching leads to a unique phase segregation behaviour, resulting in polarized states due to the local particle dynamics and not boundary effects. Hence this situation is qualitatively different from the boundary-induced phase transitions reported in Ref.~\cite{frey2lane,frey2lane1}. Moreover, cooperative lane switching induces phase separation with overall particle number conservation, unlike previously proposed  scenarios where particles moving on a lane are surrounded by a particle reservoir~\cite{lipo} or an effective reservoir~\cite{santen,santen1}. Accordingly, the transitions described in our case differ qualitatively from previously analyzed models.

In Section~\ref{sec:model} we describe the proposed theoretical framework, the rules of the dynamics and the corresponding equations of motion for each kind of species in the two lanes. In section~\ref{sec:mf} we analyze the mean field steady state solution in the continuum limit and discuss the steady states for asymmetric and symmetric exchange rate of particles between the lanes. For asymmetric exchange rates, the mean field steady state is a polarized configuration with complete particle segregation between the two lanes; with each kind of species occupying separate lanes, provided the overall density of each individual species can be accommodated in the same lane. We show how the phase segregated state arises due to the presence of an absorbing state as a steady state solution in this model. For symmetric lane exchange rates, the MF analysis allows for a family of possible steady state solutions including the phase segregated states. We analyze how  Monte Carlo studies reveal that for symmetric lane exchange rates, occurrence of a polarized steady state is a finite size effect which does not survive in the thermodynamic limit. As the system size increases, rather than attaining the absorbing state, the system remains struck in an apolar symmetric configuration. We conclude in section ~\ref{sec:conclusion} by analyzing the possible ramifications of this model for intracellular transport and discuss our results in context of other driven multi-lane transport models. 

\begin{figure}[h]
\centering
\includegraphics[width= 4in,height = 2in,angle=0]{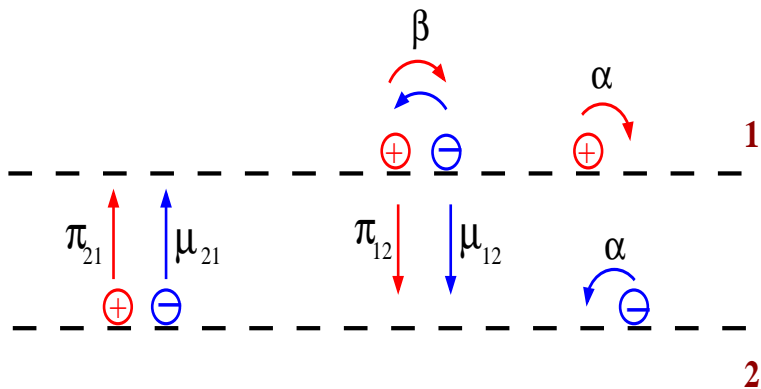}
\begin{center}
\caption{Dynamical rules for the minimal model involving translocation and switching. The lane switching processes between the two lanes are also depicted. Lane switching can occur only if the oppositely directed species are in adjacent lattice site, with their individual hopping direction facing each other.}
\label{fig1}
\end{center}
\end{figure}    
 
\section{The two lane model}
\label{sec:model}
We consider two parallel lanes, labeled  $1$ and $2$, described as two finite parallel one dimensional lattices of length $L$ with $N$ sites. These two parallel lanes mimic the biologically relevant geometry associated to   intracellular transport, with multiple parallel microtubule tracks on which cellular cargoes (vesicles and organelles) are actively carried. Therefore, the  cellular cargoes  transported on these filaments can be thought of as  effective  entities or particles, in our treatment. Consistent with the experimental evidence that cargoes are carried by molecular motors either toward  the $plus$ or the $minus$ end of the microtubule filament, we consider that the effective particles on the lattice can be either in a $(+)$- or  $(-)$-directed state.

For intracellular transport, track switching can be induced by an internal regulatory mechanism, attributed to the fact that the motors carrying the cargo experience an increased load force due to the oppositely directed cargo. Thus there is an increased propensity to detach from the filament and join the adjacent filament \cite{lipo1}. We will account for  this correlated switching mechanism, through appropriate  kinetic rules for  the effective particles which represent the transported cargoes.

The microscopic state of the effective particles moving along the lanes are characterized in terms of the occupation number at each lattice site. Specifically, $n_{i,j}^{+(-)}$ describes  a particle at site $i$ moving towards the right(left) end of lane $j$. Each site in a particular lane is occupied by either $(+), (-) $ or a vacancy $(0)$ and multiple occupancy at a given site is not allowed. A $(+)$-moving particle will jump  to the right  with the  prescribed  rate $\alpha$ if the adjacent site to the right is vacant. Similarly,  a $(-)$-moving particle will jump  to the left  with  the same rate $\alpha$ if the adjacent site to the left is vacant.  For a lattice site occupied by a $(+)$-moving particle, if the neighbouring site to the right is occupied by a $(-)$-moving particle, then two different  processes can take place;  either there is a translocation event  where  the $(+)$-moving particle hops to the neighbouring site at $i + 1$ on the same lattice while the neighbouring $(-)$-moving particle hops to the site $i$   with a rate $\beta$, or the $(+)$-moving  particle in 1(2) switches with rate $\pi_{12}(\pi_{21})$ to the corresponding site $i$ on the other lane if that site is vacant.

\begin{figure}[h]
\centering
\includegraphics[width=2.5in, angle=-90]{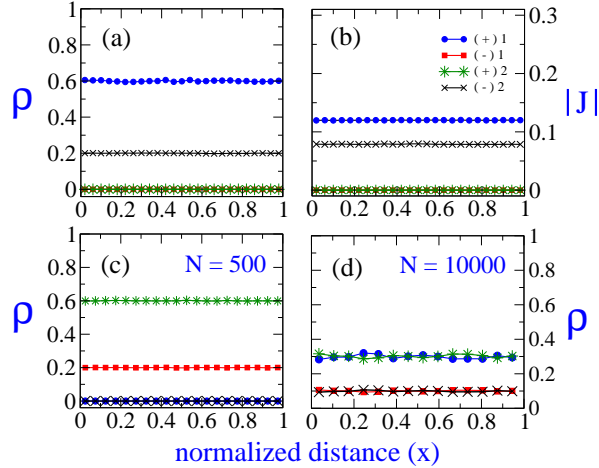}
\begin{center}
\caption{The steady state density and current profile for $(+)$ and $(-)$ for asymmetric exchange rate of particles is shown in (a) and (b), with the lane exchange rates being, $\pi_{12} = 0.2$, $\pi_{21} = 0.4$, $\mu_{12} = 0.4$, $\mu_{21} = 0.2$ and $N =5000$. This steady state is an absorbing state with complete phase segregation. (c) and (d) shows the density profile for $(+)$ and $(-)$ for the case of symmetric lanes with $ \pi_{12} = \pi_{21} = 0.4$ and $\mu_{12} = \mu_{21} = 0.4$. As shown in (c), for small system size $(N = 500)$, the system reaches the absorbing steady state while for (d) with a system size of $N =10000$, the system remains struck in state where, the density of $(+)$ in lane $1$ and $2$ are equal. Here for all the illustrated cases, the average density for  $(+)$ and $(-)$ are  $p_{av} = 0.3$ and $n_{av} = 0.1$ respectively while the translocation rates are, $\alpha = 0.5$ and $\beta = 0.3$ respectively.}
\label{fig2}
\end{center}
\end{figure}

Similar processes determine  the displacement of $(-)$-moving particles. Specifically, if the neighbouring site to the left of $(-)$-moving particle is occupied with an oppositely directed $(+)$-moving particle,  then either there is a translocation  where the $(-)$-moving  particle jumps to the neighbouring site at $i - 1$ on the same lattice while the neighbouring $(+)$ hops to the site $i$   with rate $\beta$, or  the $(-)$-moving particle switches to the other lane at the corresponding lattice site  $i$ with a rate $\mu_{12} ( \mu_{12})$ if the site $i$ of the opposite lane is vacant. All the described relevant dynamic processes are schematically depicted in Fig.\ref{fig1}.
The corresponding  kinetic equations, which  can be expressed in terms of the mean site occupation density for the two oppositely moving species in each lane,  read
\begin{eqnarray}
\frac{d\langle n_{i,1}^{+}\rangle}{dt} &=&\alpha\left \langle n_{i-1,1}^{+}(1 - n_{i,1}^{+} -n_{i,1}^{-}) - n_{i,1}^{+}(1 - n_{i+1,1}^{+} -n_{i+1,1}^{-})\right\rangle
+ \beta \langle n_{i-1,1}^{+}n_{i,1}^{-} - n_{i,1}^{+}n_{~i+1,1}^{-}\rangle \nonumber\\
&+&\pi_{21} \left\langle n_{i,2}^{+}n_{i+1,2}^{-}(1 - n_{i,1}^{+} -n_{i,1}^{-})\rangle - \pi_{12}\langle n_{i,1}^{+}n_{i+1,1}^{-}(1 - n_{i,2}^{+} -n_{i,2}^{-})\right\rangle\\
\frac{d\langle n_{i,1}^{-}\rangle}{dt} &=&\alpha\left \langle n_{i+1,1}^{-}(1 - n_{i,1}^{+} -n_{i,1}^{-}) - n_{i,1}^{-}(1 - n_{i-1,1}^{+} -n_{i-1,1}^{-})\right\rangle
+ \beta \langle n_{i+1,1}^{-}n_{i,1}^{+} - n_{i,1}^{-}n_{~i-1,1}^{+}\rangle\nonumber \\
&+&\mu_{21}\left\langle n_{i,2}^{-}n_{i-1,2}^{+}(1 - n_{i,1}^{+} -n_{i,1}^{-})\rangle -\mu_{12}\langle n_{i,1}^{-}n_{i-1,1}^{+}(1 - n_{i,2}^{+} -n_{i,2}^{-})\right\rangle\\
\frac{d\langle n_{i,2}^{+}\rangle}{dt} &=&\alpha\left \langle n_{i-1,2}^{+}(1 - n_{i,2}^{+} -n_{i,2}^{-}) - n_{i,2}^{+}(1 - n_{i+1,2}^{+} -n_{i+1,2}^{-})\right\rangle
+ \beta \langle n_{i-1,2}^{+}n_{i,2}^{-} - n_{i,2}^{+}n_{~i+1,2}^{-}\rangle \nonumber\\
&+&\pi_{12}\left \langle n_{i,1}^{+}n_{i+1,1}^{-}(1 - n_{i,2}^{+} -n_{i,2}^{-})\rangle -\pi_{21}\langle n_{i,2}^{+}n_{i+1,2}^{-}(1 - n_{i,1}^{+} -n_{i,1}^{-})\right\rangle\\
\frac{d\langle n_{i,2}^{-}\rangle}{dt} &=&\alpha\left \langle n_{i+1,2}^{-}(1 - n_{i,2}^{+} -n_{i,2}^{-}) - n_{i,2}^{-}(1 - n_{i-1,2}^{+} -n_{i-1,2}^{-})\right\rangle
+ \beta \langle n_{i+1,2}^{-}n_{i,2}^{+} - n_{i,2}^{-}n_{~i-1,2}^{+}\rangle\nonumber \\
&+&\mu_{12}\left \langle n_{i,1}^{-}n_{i-1,1}^{+}(1 - n_{i,2}^{+} -n_{i,2}^{-})\rangle -\mu_{21}\langle n_{i,2}^{-}n_{i-1,2}^{+}(1 - n_{i,1}^{+} -n_{i,1}^{-})\right\rangle
\end{eqnarray}

where $\langle n_{i,1(2)}^{\pm}\rangle$ stands  for the mean density at lattice site  $i$ in lane 1(2) of left(right) moving particles, and where the terms on the right hand side of the equations describe the corresponding  gain and loss terms arising from translocation and  lane switching processes.  
 
\section{Mean-Field equations and steady states}
\label{sec:mf}
The dynamics of oppositely moving particles introduced in the previous section is non-linear, due to the correlated switching and translocation events which determine how particles can displace along the tracks. If one neglects particle spatial correlations and factorizes the corresponding  two-point and three-point correlation function arising out of the different combination of $n_{i,j}^{+(-)}$ as product of their averages, then  
\begin{equation}
\langle n_{i,j}^{\pm}n_{i+1,j}^{\pm}\rangle = \langle n_{i,j}^{\pm}\rangle \langle n_{i+1,j}^{\pm}\rangle\\
\end{equation}

It is now possible to derive the explicit  MF equation for the evolution of the mean  particle densities, $\langle n_{i,j}^{+}\rangle = p_{j}$, $\langle n_{i,j}^{-}\rangle = n_{j}$. Here, in the notation for the mean particle density, we have not displayed the dependence on lattice position, $i$ and time, $t$ for the sake of simplicity. We are interested in situations where we have a large number of moving particles $N$, on lanes of length $L$, which are much larger than the particle size. Accordingly, we consider that the lattice spacing, $\epsilon$ vanishes in the thermodynamic limit, $\epsilon = \frac{L}{N}\rightarrow 0$, while the rescaled positions along the lanes  satisfy $x = \frac{i}{N-1}, 0 \leq x \leq 1$. The corresponding densities, which become functions of the position $x$ along the filament can be derived by performing a Taylor expansion in powers of the lattice spacing $\epsilon$. 

We will consider that all the kinetic rates which determine particle displacement and lane switching are controlled by local  interactions, and hence do not scale with lane size; hence they are of order $\epsilon^0$. In this regime, the MF  evolution equations for the mean densities in the continuum limit read
\begin{eqnarray}
\frac{\partial p_{1}}{\partial t}&=& \pi_{21} p_{2} n_{2}( 1 - p_{1} - n_{1}) -  \pi_{12} p_{1}n_{1}( 1 - p_{2} - n_{2}) -\epsilon \frac{\partial}{\partial x}\left[\alpha p_{1}( 1 - p_{1} - n_{1}) + \beta p_{1}n_{1}\right] \nonumber\\
&+& \epsilon\left[ -\pi_{12} p_{1}( 1 - p_{2} - n_{2})\left(\frac{\partial n_{1}}{\partial x}\right)+ \pi_{21} p_{2}( 1 - p_{1} - n_{1})\left(\frac{\partial n_{2}}{\partial x}\right)\right] + O(\epsilon^{2})
\label{eq:mf1} 
\\
\frac{\partial p_{2}}{\partial t}&=& \pi_{12} p_{1} n_{1}( 1 - p_{2} - n_{2}) -  \pi_{21} p_{2}n_{2}( 1 - p_{1} - n_{1}) -\epsilon \frac{\partial}{\partial x}\left[\alpha p_{2}( 1 - p_{2} - n_{2}) +\beta p_{2}n_{2}\right] \nonumber\\
&+& \epsilon\left[ -\pi_{21} p_{2}( 1 - p_{1} - n_{1})\left(\frac{\partial n_{2}}{\partial x}\right)+ \pi_{12} p_{1}( 1 - p_{2} - n_{2})\left(\frac{\partial n_{1}}{\partial x}\right)\right] + O(\epsilon^{2})
\label{eq:mf2} 
\\
\frac{\partial n_{1}}{\partial t}&=& \mu_{21} p_{2} n_{2}( 1 - p_{1} - n_{1}) -  \mu_{12} p_{1}n_{1}( 1 - p_{2} - n_{2}) +\epsilon \frac{\partial}{\partial x}\left[\alpha n_{1}( 1 - p_{1} - n_{1}) + \beta p_{1}n_{1}\right] \nonumber\\
&+& \epsilon\left[\mu_{12} n_{1}( 1 - p_{2} - n_{2})\left(\frac{\partial p_{1}}{\partial x}\right) - \mu_{21} n_{2}( 1 - p_{1} - n_{1})\left(\frac{\partial p_{2}}{\partial x}\right)\right] + O(\epsilon^{2})
\label{eq:mf3} 
\\
\frac{\partial n_{2}}{\partial t}&=& \mu_{12} p_{1} n_{1}( 1 - p_{2} - n_{2}) -  \mu_{21} p_{2}n_{2}( 1 - p_{1} - n_{1}) +\epsilon \frac{\partial}{\partial x}\left[\alpha n_{2}( 1 - p_{2} - n_{2}) + \beta p_{2}n_{2}\right] \nonumber\\
&+& \epsilon\left[ \mu_{21} n_{2}( 1 - p_{1} - n_{1})\left(\frac{\partial p_{2}}{\partial x}\right)- \mu_{12} n_{1}( 1 - p_{2} - n_{2})\left(\frac{\partial p_{1}}{\partial x}\right)\right] + O(\epsilon^{2})
\label{eq:mf4} 
\end{eqnarray}

\noindent
where  we have retained terms up to first order in $\epsilon$.\\
The corresponding MF expression for the particle current of $(+)$ and $(-)$ in lane $j$ reduce to,
\begin{eqnarray}
J_{j}^{+} &=& + \alpha p_{j}(1 - p_{j} - n_{j}) +  \beta p_{j}n_{j}\nonumber \\ 
J_{j}^{-} &=& -\alpha n_{j}(1 - p_{j} - n_{j}) - \beta p_{j}n_{j}
\label{eq:current1}
\end{eqnarray}

\begin{figure}[h]
\centering
\includegraphics[width=2.5in,height = 4.0in, angle=-90]{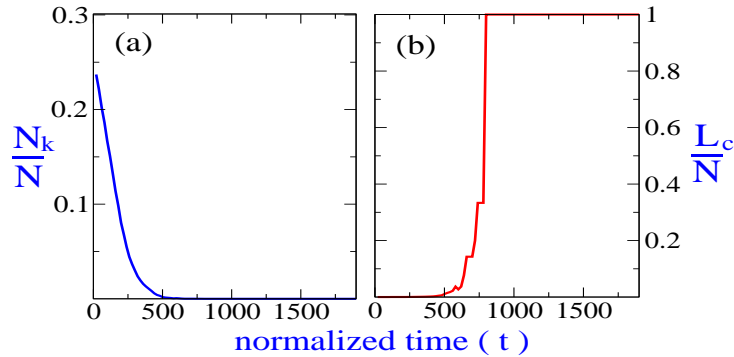}
\begin{center}
\caption{Phase separation kinetics for asymmetric lane exchange rates for a system of $N= 50000$ particles in terms of the time evolution of a) the kink number and b) the average cluster size. Here the normalized dimensionless time is in the units of $\alpha^{-1}$. (a) Eventually the number of kinks goes to zero when the system reaches the absorbing steady state. (b) The cluster size becomes equal to the system size when the absorbing state is reached. Here, $\pi_{12} = 0.15$, $\pi_{21} = 0.1$,  $\mu_{12} = 0.1$, $\mu_{21} = 0.15$, $p_{av} = 0.3$, $n_{av} = 0.2$, $\beta = 0.4$ and $\alpha= 1$.}
\label{fig3}
\end{center}
\end{figure}

The steady state configurations of oppositely moving particles must satisfy the symmetry relations
\begin{eqnarray}
\pi_{21} p_{2}n_{2}( 1 - p_{1} - n_{1}) &=& \pi_{12}p_{1}n_{1}( 1 - p_{2} - n_{2})
\label{eq:ss1}
\\
\mu_{21} p_{2}n_{2}( 1 - p_{1} - n_{1}) &=& \mu_{12}p_{1}n_{1}( 1 - p_{2} - n_{2})  
\label{eq:ss2}
\end{eqnarray}

Since we are interested in the collective behavior induced by the competing dynamics of oppositely moving particles, we will  concentrate in the simplest geometry where we disregard the role played by incoming and outgoing particles.  We leave the analysis of the effect of open boundaries on particle dynamics for  a future study and assume   periodic boundary conditions. In this case, the overall density of $(+)$ and $(-)$ is conserved, so that for homogeneous steady configurations we have,
\begin{eqnarray} 
p_{1} + p_{2} &=& 2p_{av} \nonumber\\ n_{1} + n_{2} &=& 2n_{av}
\label{eq:constant}
\end{eqnarray}
where $p_{av}$ and $n_{av}$ are the fixed total initial density of $(+)$ and $(-)$ species respectively.\\
Having obtained the steady state conditions, one can subsequently identify the allowed steady states. First we consider the case of asymmetric particle exchange between the lanes; for which the forward rate of exchange of particles from lane $1$ to $2$ is not equal to its reverse rate, {\sl i.e.}   $\pi_{21} \neq \pi_{12} $ and $\mu_{21} \neq \mu_{12}$. Subsequently, we will discuss the peculiarities of  symmetric particle lane exchange.

\begin{figure}[h]
\centering
\includegraphics[width=2.5in,height = 3.5in, angle=-90]{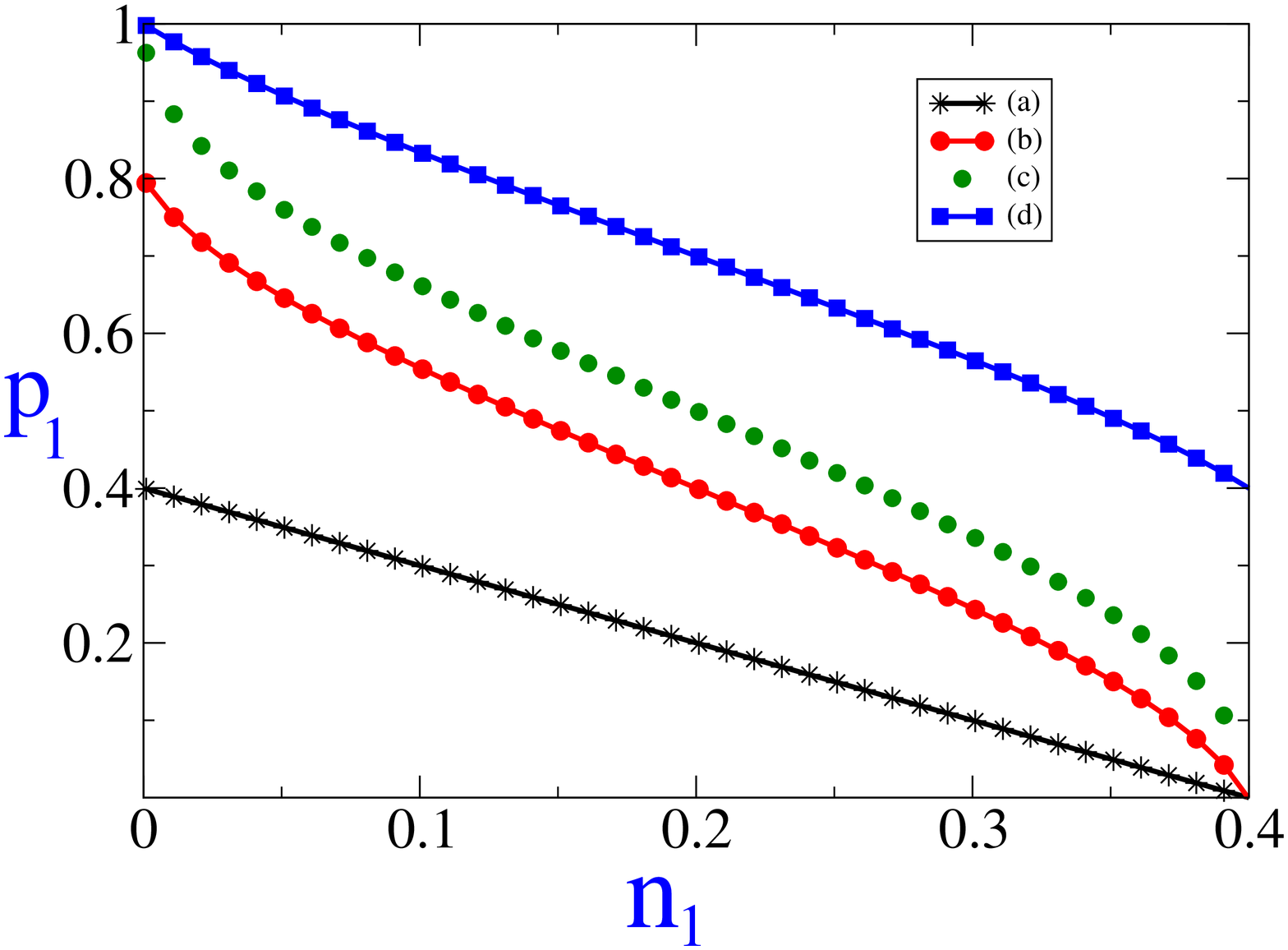}
\begin{center}
\caption{Mean Field family of steady state solution for symmetric lane exchange: Plot of $p_{1}$ vs $n_{1}$ is shown, corresponding to Eq.(\ref{eq:quadratic}). The plots are for (a) $p_{av} = 0.2$, (b) $p_{av} = 0.4$, (c) $p_{av} = 0.5$  and (d) $p_{av} = 0.7$. Here the average total density of $(-)$, $n_{av} = 0.2$ is held constant. For $p_{av}\geq 0.5$, one of the steady state corresponds to complete coverage  by $(+)$ species in one of the lanes. We note that the family of steady state includes the absorbing steady state and the steady state corresponding to equal density of each species in the two lanes.}
\label{fig4}
\end{center}
\end{figure}

\begin{figure}[h]
\centering
\includegraphics[width=2.5 in,height = 4.0in, angle=-90]{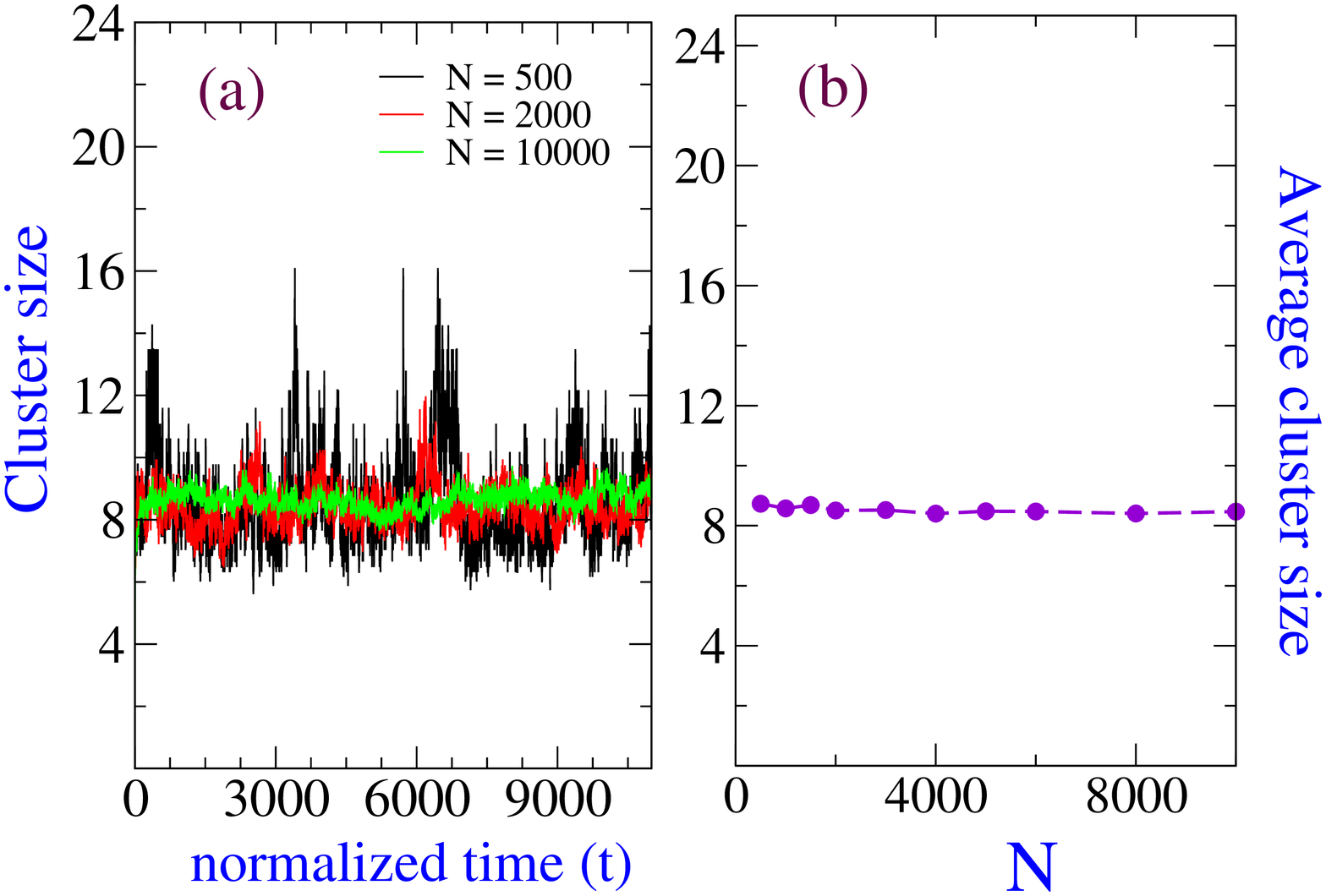}
\begin{center}
\caption{(a) Cluster size as a function of time for symmetric lane exchange rate: For symmetric exchange between the lanes, the cluster size fluctuates about a mean value with the relative fluctuations about the mean cluster size diminishing with system size $N$. (b) The average cluster size remains almost constant for variations of system size indicating the absence of phase segregation between the two lanes. Here $\alpha= 0.5$, $\beta = 0.2$, $\pi_{12} = \pi_{21} = 0.4$, $\mu_{12} = \mu_{21} = 0.4$, $p_{av} = 0.4$, $n_{av} = 0.1$. The normalized time is in the units of $\alpha^{-1}$.}
\label{fig6}
\end{center}
\end{figure}

\subsection{Asymmetric exchange between lanes}
For asymmetric lane exchange ($\pi_{21} \neq \pi_{12}$,  $\mu_{21} \neq \mu_{12}$ ), the steady state conditions,  Eqs.(\ref{eq:ss1}) and (\ref{eq:ss2}), imply  that   $p_{2}n_{2}( 1 - p_{1} - n_{1})$ and $p_{1}n_{1}( 1 - p_{2} - n_{2})$ vanish independently, unless  the lane exchange rates satisfy $\frac{\pi_{21}}{\mu_{21}} = \frac{\pi_{12}}{\mu_{12}}$. This allows for  steady state configuration in which either the two species of particle are completely phase segregated with each species occupying separate lanes or atleast one of the lanes is totally covered with one species of particle and no vacancies. As a result, a number of different steady states can develop depending on the overall number of right and left moving species.

\begin{itemize}
\item$2 p_{av} < 1, 2 n_{av} < 1$: The system phase segregates totally with the $(+)$ particles occupying  one  lane while $(-)$ particles occupy the other lane as shown in Fig.\ref{fig2}(a). If $\pi_{21} + \mu_{12} > \pi_{12} + \mu_{21}$, the homogeneous steady state solution is  $p_{1} = 2 p_{av}, p_{2} = 0, n_{1} = 0, n_{2} = 2 n_{av}$,  while in the opposite case, when $\pi_{21} + \mu_{12} < \pi_{12} + \mu_{21}$, the lane populations are  $p_{1} = 0 , p_{2} = 2 p_{av}, n_{1} = 2 n_{av}, n_{2} = 0$.
\item$2 p_{av} > 1, 2 n_{av} < 1$: For these densities one of the lanes is completely filled with $(+)$, with no vacancies so that the current in that lane is zero. The other lane contains  all the remaining $(+)$-  and all the  $(-)$-moving  particles. The selection of the lanes is again determined by whether $\pi_{21} + \mu_{12}$ is greater or less than $\pi_{12} + \mu_{21}$, like in the previous case.
\item$2 p_{av} <1, 2 n_{av} > 1$: The solution is similar to the previous situation  except that now it is the $(-)$-moving particles  which completely occupy one of the lanes with rest of the particles occupying the other lane.
\item $2 p_{av} =1, 2 n_{av} = 1$: This is a special case where both  lanes are fully occupied and therefore although there is dynamics in each of the lanes for a finite $\beta$, there is no interchange of particles between the lanes and therefore the initial starting density of individual species in each lane is separately conserved.  
\end{itemize}
All  the steady states described  are absorbing  states because once the system gets into the corresponding steady configuration, there is no particle exchange between the lanes and no microscopic site dynamics can take it out of that state.

In order to validate the  theoretical scenarios derived from the MF analysis, we have performed Monte Carlo simulations. To this end, we consider two lanes on which particles perform the rules described in Section~\ref{sec:model}. As described in Ref.\cite{ignapre}, a particle is selected at random to perform a  Monte Carlo move. A move for a  particular process (i.e; translation or lane switching) is done proportional to its rate. We start from a random initial distribution of particles and let the system evolve before collecting the data for averaging. We wait for an initial transient $\geq 1000 \frac{N}{r}$ swaps,  where $r$ stands for the rate of the slowest process ( among lane switching  or translation rates ), for the system to  attain its steady state. We then gather statistics of the relevant quantities  averaging typically  over $10^4$ time swaps and collect information with a period  $\geq 10 \frac{N}{r}$.
Monte Carlo simulations validate both the steady states density and particle currents  predicted by the MF theory as shown in  Fig.\ref{fig2}(a) and  Fig.\ref{fig2}(b) for asymmetric lane exchange with occurrence of complete phase segregation.\\
In order to assess how the system reaches the absorbing states, we have analyzed  the number of kinks ($N_{k}$), defined as the   total number of {\it sign} changes in the two lanes. We also define a closely related quantity; the average cluster size ($L_c$), identified as the average number  of consecutive set of $(+)$- or $(-)$- moving particles (which includes the count of the vacancies in between) until it encounters an oppositely directed particle. Starting from a random initial configuration, Fig.\ref{fig3} shows the fast decay of the number of kinks and the associated growth of the characteristic cluster size. The asymptotic decay of the temporal evolution of $N_{k}(t)$ is compatible with a stretched exponential form, $\exp \left[-{(\frac{t}{\tau})}^{\nu}\right]$ as shown in Fig.\ref{fig8}. The value of the exponent $\nu$ depends on the particular choice of parameter values related to translation and lane switching. During simulations, we have observed that the exponent $\nu$ can vary over a wide range, with it being even less than unity for certain range of parameter values. The stretched exponential decay of the kink number suggests a competition of relaxation time scales during kink annihilation.

\begin{figure}[h]
\centering
\includegraphics[width=2.5in,height = 4.0in, angle=-90]{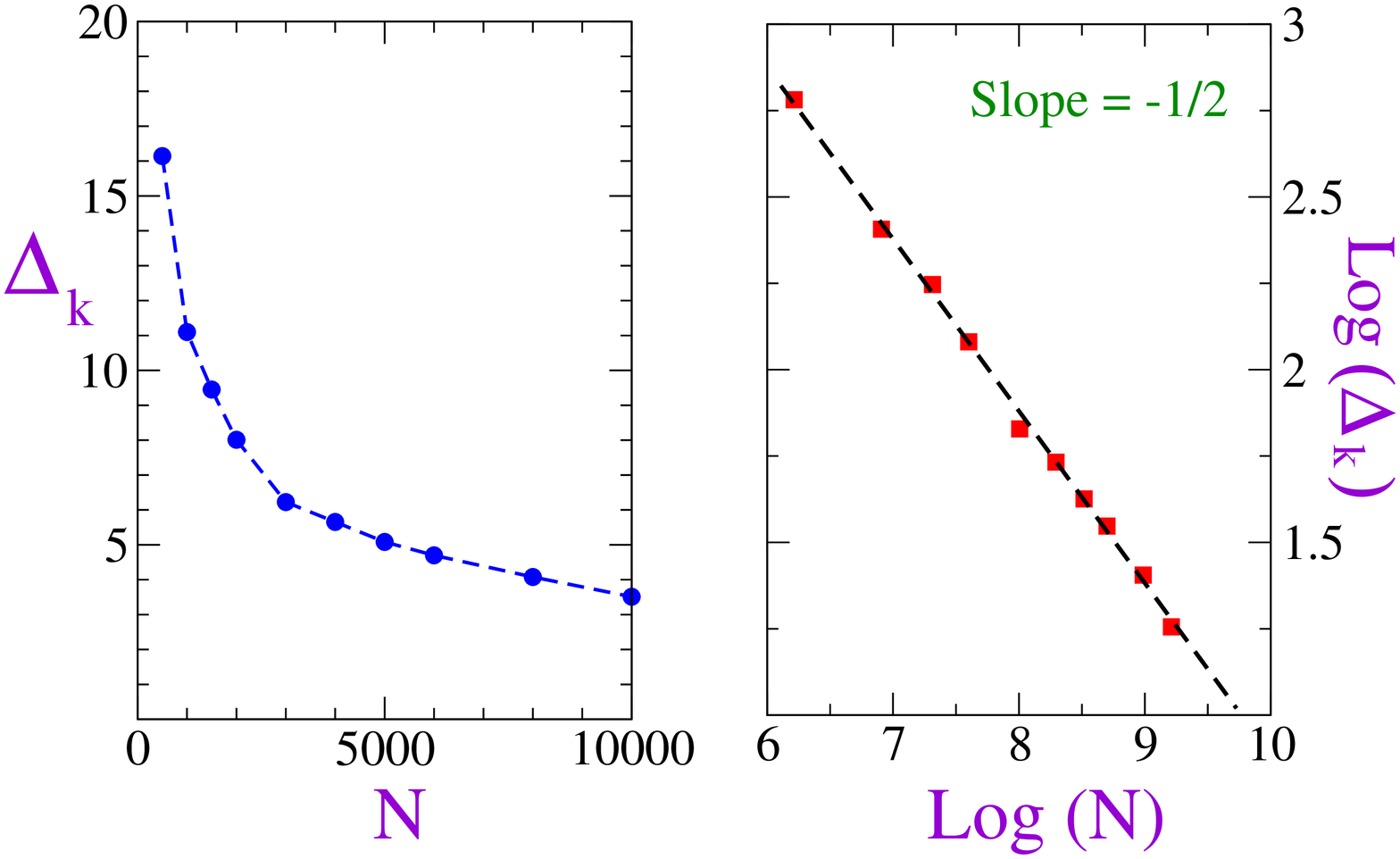}
\begin{center}
\caption{Variation  of $\Delta_k$, the relative fluctuation of the kink number, with system size. From the slope of the log-log plot which quantifies the decay of the kink number with system size, it can be inferred that $\Delta_k \sim N^{-1/2}$. So in the thermodynamic limit of $N\rightarrow \infty$, there is no phase separation and the steady state is described by equal densities of each species in both  tracks. Here $\alpha= 0.5$, $\beta = 0.2$, $\pi_{12} = \pi_{21} = 0.4$, $\mu_{12} = \mu_{21} = 0.4$, $p_{av} = 0.4$, $n_{av} = 0.1$.}
\label{fig7}
\end{center}
\end{figure}

\begin{figure}[h]
\centering
\includegraphics[width=2.5 in,height = 4.0in, angle=-90]{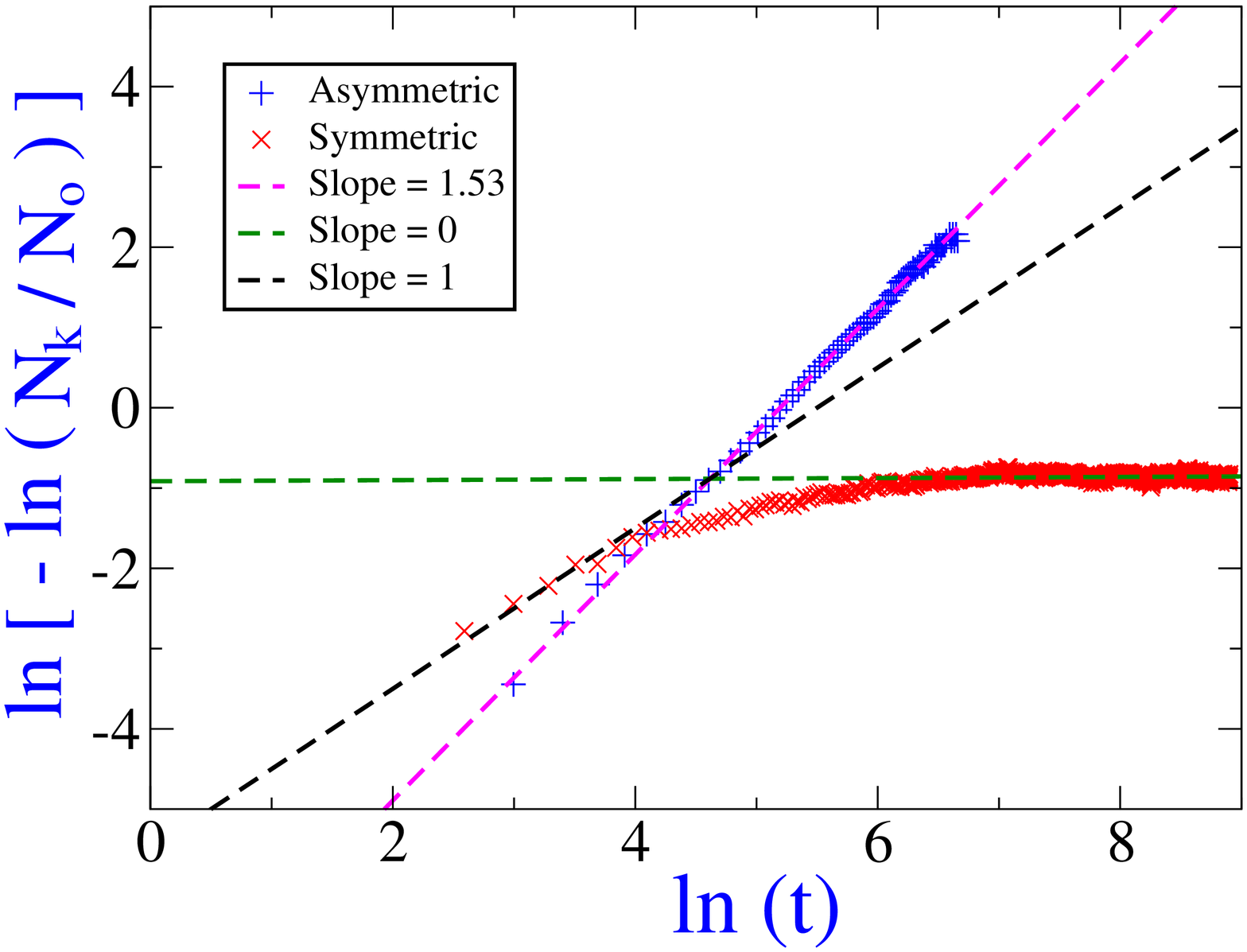}
\begin{center}
\caption{The tail of the kink number distribution is fitted to a stretched exponential form, $N_{k} \sim N_{o}\exp\left[-{(\frac{t}{\tau})}^{\nu}\right]$. For the case of asymmetric lane exchange, the slope $\nu = 1.53$ corresponds to a particular choice of parameter values, $\pi_{12}= 0.15, \pi_{21} = 0.1$, $\mu_{12}= 0.1$, $\mu_{21} = 0.15$, $\alpha= 1$, $\beta = 0.4$, $p_{av} = 0.3$ and $n_{av} = 0.2$. For symmetric lane exchange, the slope is always zero and thus indicating that the average kink number remains constant. For the symmetric case, the data points are plotted for $\pi_{12}= \pi_{21} = 0.15$, $\mu_{12}= \mu_{21} = 0.15$. A line with slope 1, corresponding to an exponential decay with time is shown for comparison with the actual data points. Time is in the units of $\alpha^{-1}$ and the system size $N = 50000$.}
\label{fig8}
\end{center}
\end{figure}

\subsection{Symmetric exchange between lanes}
For symmetric exchange rates between the two lanes, when  $\pi_{21} = \pi_{12} = \pi$ and $\mu_{21} = \mu_{12} = \mu$,  the steady states must fulfill the symmetry relation
\begin{equation}
p_{2}n_{2}( 1 - p_{1} - n_{1}) = p_{1}n_{1}( 1 - p_{2} - n_{2})
\label {eq:ss-sym}
\end{equation}

Using the conservation laws in Eq.(\ref{eq:constant}), the particle densities in each of the lanes follow a quadratic equation, 
\begin{equation}
(n_{av} -n_{1})p_{1}^{2} + \left[2n_{1}(p_{av} + n_{av}) -2 p_{av}n_{av} -n_{av} -n_{1}^{2}\right]p_{1} + (1-n_{1})(2 p_{av}n_{av} - p_{av}n_{1}) = 0
\label{eq:quadratic} 
\end{equation} 

which implies that as opposed to the case of asymmetric exchange, the MF analysis does not lead to a unique steady configuration. Rather, it predicts a family of allowed steady states. Fig.~\ref{fig4}, which  displays the allowed configurations on varying the overall fraction of $(+)$-moving particles, illustrates that both, the  completely segregated state ($p_{1} = 0, n_{1} = 2 n_{av}$ or $n_{1} = 0, p_{1} = 2 p_{av}$) and state with equal density in both lanes are allowed MF solution. Although a linear stability analysis of the different steady state configurations, discussed in Appendix A, does not resolve this degeneracy, Monte Carlo simulations  clearly indicate  a unique steady state. However, the simulations indicate a strong sensitivity of the system kinetics to the system size.  Starting from an analogous initial configuration, Fig.~\ref{fig2}(c) shows that for a small lane the steady configuration is the absorbing state, while for a larger system size, the system remains in the symmetric configuration, where  $p_{1} = p_{2} = p_{av}, n_{1} = n_{2} = n_{av}$, corresponding to equal density of particles in both lanes, as shown in Fig.~\ref{fig2}(d).

We have performed a more systematic study to assess the impact of the system size on the kinetics of  particle motion. Fig.~\ref{fig6} shows that at large system sizes, the fluctuations in  the average cluster size around the symmetric configuration decrease with system size. This feature can be further understood by analyzing the  relative fluctuations in the number of kinks $\Delta_k$, as a function of system size, as depicted in Fig.~\ref{fig7}. There is a clear algebraic decay in the amplitude of the fluctuations of the kink number ($\Delta_k$ ), as total particle number $N$ increases, compatible with  $\Delta_k\sim 1/N^{-1/2}$. This algebraic decay indicates that if driven kinetically to a symmetric configuration, as we observe in all the Monte Carlo runs carried out, the time required by the system for escaping to the allowed polarized lane occupation, will increase exponentially with system size. In the  thermodynamic limit, $N\rightarrow\infty$, the system would be trapped kinetically in this symmetric configuration.

Therefore, the symmetric  lane exchange leads to a qualitatively different scenario from their asymmetric counterparts.
 In the former case, the degeneracy  in the allowed steady state configurations allow for competition between different  configurations so that, the oppositely moving particles can be kinetically captured in a symmetric configuration. The escape kinetics is no longer controlled by the lane  exchange rate; rather one observes a slow dynamics analogous to the one observed in critical phenomena. These numerical evidence suggests that in the thermodynamic limit of $N\rightarrow \infty$, the system is never able to attain the absorbing configuration and instead the actual steady state corresponds to equal density of each species in both lanes. Fig.~\ref{fig8} illustrates how the relaxation dynamics of the kink numbers for the symmetric case is distinct from the asymmetric case; while the kink numbers decay to zero for the asymmetric case, the average number of kinks remains roughly constant after the initial transient for the symmetric case. 

\section{Summary and Conclusions}
\label{sec:conclusion}
We have presented a two lane model for transport with oppositely directed particles and correlated lane switching between the lanes. This system has been analyzed via Mean field theory and Monte Carlo simulations. Owing to the correlated lane switching, the system exhibits  absorbing state configurations as an allowed stationary state.  This absorbing state corresponds to a situation of phase segregation of particles between the lanes resulting in a polarized state. For asymmetric particle exchange between the lanes, MF analysis predicts a unique steady state corresponding to such an absorbing state. When the average density of both  $(+)$- and $(-)$- moving particles is smaller than $1/2$, then the system undergoes complete polarization, with oppositely directed particles occupying separate lanes. However, for symmetric particle exchange rates, a system starting from a random initial condition is driven kinetically through a symmetricexchupat leastato a om witat the time scale to escape grows exponentially with system size. As a result, phase segregation has only been observed numerically for relatively small system sizes. Therefore, in the thermodynamic limit of $N\rightarrow\infty$, the system never reaches this absorbing steady state and it remains stuck in a symmetric steady state characterized by an equal distribution of particles in both the lanes. \\
The temporal relaxation dynamics of the system to the absorbing steady state is studied in terms of the time evolution of the kink number $N_{k}$ and cluster size. For  asymmetric lane exchange, after an initial transient, the time evolution of the kink number follows a stretched exponential form, indicating that several time scales control the system relaxation  to  the absorbing steady state. It would be interesting to conduct future studies to probe the underlying coarsening mechanism in  such driven multi-lane systems. \\
The steady state behaviour and phase segregation described differs qualitatively from  two species driven systems with  cooperative Langmuir (un)binding processes, which are in contact with particle reservoir \cite{lipo}. Although non-equilibrium phase transitions have been described in these systems, they lack  absorbing steady states. Moreover, in the systems described  here,  the observed phase segregation resulting in polarized state arises due to local correlated switching dynamics between the lanes and not due to boundary effects like in case of systems studied in \cite{frey2lane, pronina1}.\\
One possible extension of the present work  would be to investigate this model for the case of open boundaries, with incoming and outgoing particle flux at the boundaries. The possible competition between the boundary input/output processes with correlated switching may then open up the possibility of discovering new phases arising out of the interplay of these two distinct processes. It remains to be seen whether in that case, the system's behavior  differs markedly  from the behavior described in Ref.~ \cite{juhasz1,juhasz2}, where systems with boundary input of particles and uncorrelated switching between lanes was analyzed and shock localization in the bulk has been predicted.\\
Suitable extensions of this model would be useful in the  context of organelle  intracellular transport along microtubules where oppositely directed particles are involved. For such cases, the correlated switching between the lanes maybe be one the possible mechanisms by which  jamming is avoided and transport is regulated by local mechanisms associated with motor binding kinetics. 

\section{Acknowledgments}
SM would like to thank S. M. Bhattacharjee for useful discussions and suggestions. IP acknowledges financial support from MICINN  (Spain) and Generalitat de Catalunya  under projects  FIS2008-04386 and 2009SGR-634, respectively. 

\appendix
\section{Linear stability analysis for symmetric lanes}

If we consider the MF evolution equations; Eq.(\ref{eq:mf1}-\ref{eq:mf4}) for symmetric lanes, then in the thermodynamic limit of $\epsilon \rightarrow 0$, the equations of the homogeneous fluctuations takes the form,  
\begin{equation}
\frac{\partial}{\partial t}\left[ \begin{array}{c} \delta p_{1} \\ \delta n_{1} \end{array} \right]  = \begin{bmatrix} 2\pi G & 2\pi H \\ 2\mu G & 2\mu H \end{bmatrix}\left[ \begin{array}{c} \delta p_{1} \\ \delta n_{1} \end{array} \right]
\label{eq:fluc}
\end{equation}
where, 
\begin{eqnarray}
G &=& 2 ( n_{av} - n_{1})p_{1} + 2n_{1}(n_{av} + p_{av}) - 2 n_{av}p_{av} - n_{1}^{2} - n_{av}\nonumber\\
H &=& 2 ( p_{av} - p_{1})n_{1} + 2p_{1}(n_{av} + p_{av}) - 2 n_{av}p_{av} - p_{1}^{2} - p_{av}
\end{eqnarray}
Here $p_{1},n_{1}$ corresponds to the fixed points about which fluctuations are evaluated.\\
In order to derive Eq.(\ref{eq:fluc}), we have used the fact that for homogeneous fluctuations, $\delta p_{1} = -\delta p_{2}$ and $\delta n_{1} = -\delta n_{2}$.\\
The eigenvalues corresponding to these set of equations of fluctuations are,
\begin{eqnarray}
\lambda_{1} &=& 0 \nonumber\\
\lambda_{2} &=& 2(H\mu + G\pi)
\label{eq:eigen}
\end{eqnarray}  
The corresponding eigenvectors are, $\left[ \begin{array}{c} -\frac{H}{G} \\ 1 \end{array} \right]$  and  $\left[ \begin{array}{c} \frac{\pi}{\mu} \\ 1 \end{array} \right]$.\\
We find that the eigenvalue ($\lambda_2$) is negative in physical range of densities, for the entire set of fixed points corresponding to the family of steady state solutions Eq.(\ref{eq:quadratic}). So linear stability analysis of the MF equations does not help in finding out the actual fixed point observed in the case of Monte Carlo simulations. For example, corresponding to the fixed point $p_{1} = 2 p_{av}, p_{2}= 0, n_{1} = 0, n_{2} = 2n_{av}$ (for $p_{av} < 0.5$ and $p_{av} < 0.5$), the expression for $\lambda_2$ is, 
\begin{equation}
\lambda_{2} = 2\mu p_{av}(2n_{av} - 1 ) +  2\pi n_{av}(2p_{av} - 1 ) \nonumber
\end{equation} 		
which is negative. Similarly for the fixed point corresponding to equal distribution of particles between the lanes ($p_{1} = p_{2} = p_{av}, n_{1} = n_{2} = n_{av}$),
\begin{equation}
\lambda_{2} = 2\mu p_{av}(p_{av} - 1 ) +  2\pi n_{av}(n_{av} - 1 ) \nonumber
\end{equation} 	
which is also negative for physical parameter range.

\end{document}